\documentclass[12pt]{article}

\usepackage{amssymb}
\usepackage{amscd}
\usepackage{graphics}
\usepackage{amsfonts}
\usepackage{amsmath}
\usepackage{latexsym}
\usepackage[all]{xy}
\usepackage{fancyhdr}
\usepackage{color}
\usepackage{empheq}

\usepackage{amsthm}

\usepackage{latexsym,graphics,color}
\def\bes{\begin{subequations}}
\def\ees{\end{subequations}}
\def\ba{\begin{align}}
\def\ea{\end{align}}

\def\w{\wedge}
\def\th{\theta}

\def\re{\color{red}}
\def\be{\begin{equation}}
\def\ee{\end{equation}}
\def\ti{\tilde}

\def\br{\mathbb R}
\def\R{\mathcal R}

\def\s1{\sigma^1}
\def\s2{\sigma^2}
\def\s3{\sigma^3}

\def\si{\sigma}

\def\d{\partial}
\def\jp{\frac{1}{2}}
\def\ri{{\mathrm i}}
\def\bos{\boldsymbol}

\definecolor{lila}{rgb}{1,0.2,0.9}
\definecolor{brown}{rgb}{0.5,0.3,0.3}
\definecolor{turquoise}{rgb}{0.2,0.9,0.7}
\definecolor{Orange}{rgb}{0.93,0.44,0}           
\definecolor{GrayBlue}{rgb}{0.35,0.4,0.62}       
\definecolor{SeafoamGreen}{rgb}{0.54,0.71,0.50}  

\definecolor{darkorange}{cmyk}{.20,.50,.80,0}
\definecolor{lightorange}{cmyk}{.07,.37,.65,0}
\definecolor{darkpeagreen}{cmyk}{.50,.30,.50,0}
\definecolor{lightpeagreen}{cmyk}{.22,.20,.40,0}

\newtheorem{thm}{Veta}[section]

\theoremstyle{definition}

\theoremstyle{definition}
\newtheorem{rem}[thm]{Remark}

\theoremstyle{definition}

\def\bo{\boldsymbol 0}%
\def\ri{{\mathrm{i}}}                   %
\def\cR{{\cal R}}                       %
\def\1{{\mbox{\boldmath $1$}}}          %
\def\e{\epsilon}  %

\def\bt{\beta}                          %
\def\jp{\frac{1}{2}}                    %
\def\om{\omega}                         %
\def\Om{\Omega}                         %
\def\al{\alpha}                         %
\def\ga{\gamma}                         %

\def\an{\boldsymbol{\{}}%
\def\b{\boldsymbol{\}}}

\def\bl{\color{blue}}      
      
\definecolor{spec}{rgb}{0.0, 0.26, 0.15}

\def\bx{\boldsymbol{x}}
\def\bB{\boldsymbol{B}}
\def\bs{\boldsymbol{s}}
\def\bm{\boldsymbol{m}}
\def\bl{\boldsymbol{l}}
\def\bk{\boldsymbol{k}}

\def\bp{\boldsymbol{p}}

\def\bA{\boldsymbol A}
\def\bn{\boldsymbol n}

\def\vv{\vert}
\def\vbB{\vert\bB\vert}
\def\vbx{\vert\bx\vert}

\def\bpm{\begin{pmatrix}}
\def\epm{\end{pmatrix}}

\DeclareMathSymbol{\Rho}{\mathalpha}{operators}{"50}

\begin{document}

\begin{flushright}
{}~
  
\end{flushright}

\vspace{1cm}
\begin{center}
{\large \bf  Superintegrability, symmetry and point particle T-duality}

\vspace{1cm}

{\small
{\bf Ctirad Klim\v{c}\'{\i}k}
\\
Aix Marseille Universit\'e, CNRS, Centrale Marseille\\ I2M, UMR 7373\\ 13453 Marseille, France}
\end{center}

\vspace{0.5 cm}

\centerline{\bf Abstract}
\vspace{0.5 cm}
\noindent   We show that the ideas  related to integrability  and symmetry  play an important role not only in the string T-duality story but also in its point particle counterpart. Applying those ideas, we find  that the  T-duality seems to be a more widespread phenomenon in the context of the point particle dynamics  than it is in the string one; moreover, it concerns physically very relevant point particle dynamical systems and not just  somewhat exotic ones fabricated for the purpose.    As a source of T-duality examples, we consider maximally superintegrable spherically symmetric electro-gravitational backgrounds in $n$ dimensions.
We then describe in detail four such  spherically symmetric dynamical systems  which are all mutually interconnected by a web of point particle T-dualities. 
In particular, the dynamics of a charged particle scattered by a repulsive Coulomb potential in a flat space is T-dual to the dynamics of the Coulomb scattering in the space of constant negative curvature, but it is also T-dual to the (conformal) Calogero-Moser   inverse square dynamics both in flat and hyperbolic spaces. Thus knowing just the   Hamiltonian dynamics of the scattered particle cannot give us an information about the curvature of the space in which the particle moves.

  \vspace{2pc}
  \section{Introduction}

  The motion of a classical string in a gravitational-Kalb-Ramond background is characterized by a dynamical system referred to as a
  nonlinear $\si$-model in $1+1$ spacetime dimensions. In
  the case of a topologically trivial Kalb-Ramond field strength, the classical action of this $\si$-model reads
\be S=\int d\tau \oint d\sigma (g_{ij}(x)+b_{ij}(x))\partial_+x^i\partial_-x^j,\quad \d_\pm:=\d_\tau\pm\d_\si,\label{si}\ee
where
$\tau$, $\si$ are respectively time and (circular) space coordinates on the worldsheet, $x^i$ are coordinates on the target space $T$ and $(g_{ij}(x),b_{ij}(x))$ are a metric tensor and a Kalb-Ramond potential in those coordinates.

\medskip

Consider some other  gravitational-Kalb-Ramond background
$(\ti T,\ti g_{ij}(\ti x),\ti b_{ij}(\ti x))$ and the $\si$-model which corresponds to it
\be \tilde S=\int d\tau \oint d\sigma (\ti g_{ij}(\ti x)+\ti b_{ij}(\ti x))\partial_+\ti x^i\partial_-\ti x^j.\label{tsi} \ee
The phenomenon of stringy T-duality \cite{KY,SSe,DQ93,KS95,KS96b,KS97}
takes place if the backgrounds $(T,g ,b)$ and $(\ti T,\ti g  ,\ti b )$ are not {\it geometrically equivalent} but the
$\si$-models \eqref{si} and \eqref{tsi} are {\it dynamically equivalent}.

\medskip

The geometrical (non)equivalence of the targets
means the (non)existence of a diffeomorphism $D:T\to\ti T$ such that
$(D^*\tilde g,D^*\ti b)=(g,b)$.
On the other hand,
  the dynamical equivalence of the
$\si$-models \eqref{si} and \eqref{tsi} means the equivalence of their Hamiltonian dynamics    \cite{AAL,S97}. Thus if
the $\si$-models
\eqref{si},\eqref{tsi} are characterized by their respective
phase spaces $P$,$\ti P$, symplectic forms
$\om$, $\ti\om$ and Hamiltonians
$h$, $\ti h$ they are dynamically equivalent if it exists a symplectomorphism
$\Upsilon:P\to \ti P$ such that
$\Upsilon^*\ti h=h$.  

\medskip

In particular,
the phase spaces $P$ and $\ti P$ of the $\si$-models \eqref{si} and \eqref{tsi} are parametrized respectively by the   functions $x^i(\si),p_i(\si)$ and $\ti x^i(\si),\ti p_i(\si)$,
the symplectic forms are the canonical ones
$$\omega=\oint d\si d p_i(\sigma)\wedge  d x^i(\sigma), \quad \ti\omega=\oint d\si d \ti p_i(\sigma)\wedge  d \ti x^i(\sigma)$$ and the Hamiltonians read
$$ h(x,p)=\jp \oint d\si g^{ij}(x)\left(p_i-b_{ik}(x)\d_\si x^k \right)\left(p_j-b_{jl}(x)\d_\si x^l \right) + \jp \oint d\si g_{ij}(x)\d_\si x^i\d_\si x^j,$$
$$  \ti h(\ti x,\ti p)=\jp \oint d\si \ti g^{ij}(\ti x)\left(\ti p_i-\ti b_{ik}(\ti x)\d_\si \ti x^k \right)\left(\ti p_j-\ti b_{jl}(\ti x)\d_\si \ti x^l \right) + \jp \oint d\si \ti g_{ij}(\ti x)\d_\si \ti x^i\d_\si \ti x^j.$$
 
The symplectomorhism
$\Upsilon$ is   a canonical transformation 
$\ti x=\ti x(x,p),\ti p=\ti p(x,p)$ such that
$$ h(x,p)=\ti h(\ti x(x,p),\ti p(x,p)).$$

\bigskip

Historically, the dynamical equivalence of strings moving in the geometrically non-equivalent backgrounds came as a surprise and it  was often considered to be  a distinctive feature of the string dynamics with respect to the point particle one.  However, as it was pointed out  in \cite{K20},
the T-duality exists also in the point particle context, where it establishes the dynamical equivalence of  geometrically non-equivalent electro-magnetic-gravitational backgrounds. 

The motion of a classical point particle in a electro-magnetic-gravitational background is characterized by a classical action of a $0+1$-dimensional $\si$ model
\be S= \int dt \left( \jp g_{ij}(x) \dot x^i\dot x^j -A_i(x) \dot x^i-V(x)\right),\label{pp}\ee
where
$t$ is the time, $x^i$ are coordinates on the target space $T$ and $(g_{ij}(x),A_{i}(x), V(x))$ are respectively the metric tensor as well as the vector and the scalar potentials.  

\medskip

Consider a geometrically non-equivalent background
$(\ti T,\ti g_{ij}(\ti x), \ti A_{i}(\ti x), \ti V(\ti x))$ and the corresponding action
\be \ti S= \int dt \left( \jp \ti g_{ij}(\ti x) \dot {\ti x}^i\dot {\ti x}^j -\ti A_i(\ti x) \dot {\ti x}^i-V(\ti x)\right).\label{tpp}
\ee
 The phase spaces $P$ and $\ti P$ of the $0+1$-dimensional $\si$-models \eqref{pp} and \eqref{tpp} are parametrized respectively by the canonically conjugated coordinates $x^i ,p_i $ and $\ti x^i ,\ti p_i $,
the symplectic forms are the canonical ones
$$ \omega=  d p_i \wedge  d x^i , \quad \ti\omega=  d \ti p_i \wedge  d \ti x^i $$ and the Hamiltonians read
\be   h(x,p)=\jp g^{ij}(x) \left(p_i+A_i(x)\right) \left(p_j+A_j(x)\right)+V(x),\label{hamp}\ee
\be \tilde h(\ti x,\ti p)=\jp \tilde g^{ij}(\ti x) \left(\tilde p_i+\tilde A_i(\ti x)\right) \left(\tilde p_j+\tilde A_j(\ti x)\right)+\tilde V(\ti x).\label{thamp}\ee

In full analogy with the string case, we declare the point particle models \eqref{pp} and \eqref{tpp} mutually T-dual  if it exists a canonical transformation $\tilde x=\tilde x(x,p)$, $\tilde p=\tilde p(x,p)$ such that
$$\tilde h(\tilde x(x,p),\tilde p(x,p))=h(x,p).$$
 
 \medskip
 
The first nontrivial examples of the point particle T-duality obtained in \cite{K20} showed that the phenomenon did exist but otherwise they were not particularly physically relevant and they were fabricated for the purpose by essentially a trial and error method.
In this paper, we do much better, we show that the point particle T-duality concerns physically very relevant dynamical systems and we give also a method how to obtain many new examples. This method is based on the concepts of integrability and symmetry and was largely inspired by the string T-duality story where,  apparently, all known integrable $\si$-models are (Poisson-Lie) symmetric and T-dualizable. It turns out that the integrability and symmetry help to find the T-duality examples also in the point particle context, moreover, the reason why they help turns out to be much clearer than in the string case where the observed relation between the integrability and T-dualizability remains somewhat mysterious.

  \medskip
  
 Few remarks are perhaps in order about the motivations to study the point particle T-duality. First of all, it is an interesting problem to deal with on its own, since it opens a problem of classification of physical dynamical systems in T-duality equivalence classes. All  members of a given class share 
 the same dynamical properties, which maybe manifest or hidden depending on which representative of the class we consider. For example, the T-duality 
 between the Coulomb and the Calogero-Moser scattering, which we establish in the present paper,   means that the manifest conformal symmetry of the Calogero-Moser model is
 present also in the Coulomb one albeit in a hidden dynamical way.
 
 \medskip 
 
 Another motivation has to do with the problem of zero modes in the string T-duality story. 
   At a  first sight it might seem that at least some examples of the point particle  T-duality   could be obtained by a sort of dimensional reduction of the stringy T-duality, or, said in other words, by restricting the string dynamics to the zero modes. 
  However, this is not the case because
 (with a notable exception of Abelian T-duality)  the T-duality phenomenon in string theory was so far   established only for  strings deprived of the zero modes. Indeed, the stringy T-duality   is in reality  {\it dismembered},   that is, it takes place only if we cut out some zero modes from the string on both original and dual side. This means, in particular, that
no examples of point particle T-duality can be obtained by a dimensional reduction of this dismembered string T-duality. However, it might be possible to go in an opposite direction, this is to say, to work out viable examples of the point particle T-duality and to "glue" them
to the dismembered stringy T-duality examples to achieve a full-fledged string T-duality.

\medskip

The plan of the paper is as follows.
In Section 2, we construct a particularly simple dynamical system in $n$ dimensions that we
call {\it referential spherically symmetric maximally superintegrable  system}. Although this simple system does not have a geometric interpretation as a 
$0+1$-dimensional $\si$-model, still it plays an important role
in our analysis because
it does naturally represent a T-duality class of $0+1$-dimensional $\si$-models which do have the geometric interpretation. Indeed,
in Section 3, 
we show that four physically relevant and geometrically distinct spherically symmetric $\si$-models are 
maximally superintegrable and symplectomorphic to the referential system. 
It follows, that they are all mutually T-dual, or, said differently, they belong all to the same T-duality class represented by the referential model.   Those four systems  are  the (repulsive) Coulomb potential in the flat space and  in the space of constant negative curvature,
as well as the   Calogero-Moser potential 
in the flat and in the hyperbolic spaces.
In Section 4, we provide conclusions and an outlook.   Two technical results concerning Section 3  are placed into Appendix.
 
\section{Referential maximally superintegrable system}
  A dynamical system $(P,\omega,h)$ is a smooth manifold $P$ equipped with a symplectic form $\omega$ and with a smooth function $h$, such that all time evolution flows generated by the Hamiltonian $h$ are complete, that is, they can be all smoothly prolonged
to both forward and backward infinities $t\to \pm\infty$.

\medskip 

 Let $H>0$, $T$ be canonically conjugated coordinates on an open symplectic half-plane $P_1$ equipped with the Darboux symplectic form
 $$\omega_1=dH\wedge dT,$$
  or, equivalently,
  with the Darboux Poisson bracket
  \be \{T,H\}=1.\label{pth}\ee Note that a choice of the Hamiltonian
  $h_1(H,T)=H$ gives a honest dynamical system $(P_1,\omega_1,h_1=H)$ with the complete flows $H=$const, $T=t-t_0$. Indeed, this simple form of the flows follows from the Hamiltonian equations of motion which take the form
   $$ \dot T=\{T,h_1\}=1,\quad \dot H=\{H,h_1\}=0.$$ 
    
    \begin{rem}{\footnotesize 
    On the other hand,   a choice $h_1(H,T)=T$ does not give a
  dynamical system because the corresponding flows $T=$const,  $H=-t+t_0$ cannot be prolonged to $t\to\infty$ ($H$ must remain positive).}
      
    \end{rem}

  \medskip 
  
  Let $S^{n-1}$ be the standard $(n-1)$-dimensional unit sphere and
  $T^*S^{n-1}$ its cotangent bundle equipped
  with its standard symplectic form
  $\om_{T^*S^{n-1}}$. 
  We parametrize   $T^*S^{n-1}$ by 
  $n$-vectors $\bB$ and $\bk$ fulfilling
 \be \quad \bk\bk=1,\quad \bB\bk=0.\label{co}\ee
  The vector $\bk$ thus represents a point on the   sphere $S^{n-1}$, while $\bB$ parametrizes the cotangent space at $\bk$. The symplectic form $\om_{T^*S^{n-1}}$ then reads 
  $$\om_{T^*S^{n-1}}=d{\bos B}\w d{\bos k}.$$
  
  \medskip 
  
  We are now ready to define 
  the {\it referential spherically symmetric maximally superintegrable  dynamical system} $(P_n,\om_n,h_n)$ alluded to in the Introduction. The phase space  $P_n$ of this dynamical system is defined as 
 $$ P_n=P_1\times T^*S^{n-1},$$
  its symplectic form $\omega_n$ is given by
  \be \omega_n=\omega_1+\om_{T^*S^{n-1}}=dH\w dT+d{\bos B}\w d{\bos k}.\label{omn}\ee 
  and its Hamiltonian $h_n$ is given simply  by
  $$h_n=H. $$
  We now provide an $n$-dimensional analogue of \eqref{pth}, that is  the complete set of Poisson brackets corresponding to (or characterizing)  the symplectic form $\om_n$:
  \be \{T,H\}=1,\quad 
 \{H,\bk\}=\boldsymbol{0},\quad \{T,\bk\}=\boldsymbol{0},\quad 
 \{H,\bB\}=\boldsymbol{0},\quad \{T,\bB\}=\boldsymbol{0},\label{sep}\ee \be\{B_i,B_j\}=B_ik_j-B_jk_i,\quad
\{k_i,B_j\}=\delta_{ij}-k_ik_j,\quad \{k_i,k_j\}=0, \quad i,j=1,...,n.\label{dirb}\ee
Note that the brackets \eqref{dirb} are the Dirac ones; they are derived from \eqref{omn} by taking into account the constraints \eqref{co}.

\medskip

 The Hamiltonian equations of motion
 of the referential system $(P_n,\om_n,h_n)$ read
 $$ \dot\bk=\{\bk,h_n\}=0,\quad  \dot\bB=\{\bB,h_n\}=0,\quad\dot T=\{T,h_n\}=1,\quad \dot H=\{H,h_n\}=0$$
  and this implies
   the completeness of the
  flows $\bk$={\bf const}, $\bB$={\bf const}, $H$=const, $T=t-t_0$.  
  
  \medskip 
  
 The coordinates $\bk,\bB$ on $T^*S^{n-1}$   Poisson commute with the Hamiltonian $h_n=H$, they are therefore the integrals of motion. Together with the Hamiltonian $h_n$, those coordinates furnish the $2n-1$ integrals of motion in involution, the   referential system $(P_n,\om_n,h_n)$
 is therefore   {\it maximally superintegrable}. Note, in particular,
 that the components of the wedge product $\bk\wedge \bB$ are conserved generators of $n$-dimensional rotations (i.e. the angular momenta).
  
  \medskip

 \medskip 
 
 \begin{rem}{\footnotesize
 We note that the referential dynamical system $(P_n,\om_n,h_n)$ does not lend itself to a geometric interpretation. However, as we shall see in the next section, it is symplectomorphic to at least four dynamical systems  which do have the geometrical interpretation as the $0+1$-dimensional
 $\si$-models. We may therefore say that the non-geometric referential system naturally represents a whole T-duality equivalence class of the geometric systems.}
 \end{rem}
 
 We now add some technical stuff which will be useful in the next section. Consider a $2n$-dimensional manifold $$ M_n=\{(\bp,\bx)\in\br^n
\times\br^n,\bx\neq\boldsymbol{0}\}$$
equipped with the Darboux symplectic form
$$ \Om_n=d \bp\wedge d\bx.$$
The canonical Poisson brackets   corresponding to $\Om_n$ are
$$ \an x_i,x_j\boldsymbol{\}} =0,\quad \an p_i,p_j\b =0,\quad 
  \an x_i,p_j\b=\delta_{ij}, \quad i,j=1,\dots, n.$$
  Note that we use the notation $\{.,.\}$ for the Poisson brackets on
  $P_n$ and the boldface one $\an .,.\b$ for the Poisson brackets on $M_n$.
  
  \medskip 
  
  It turns out that the symplectic manifold $M_n$ is the phase space $P_n$ 
  in a disguise. Indeed,  consider a bijection
  $\cR^r:M_n\to P_n$ defined as 
$$ H=h^r(\bx,\bp):=\jp\bx^2,\quad T=t^r(\bx,\bp):=-\frac{\bp\bx}{\bx^2},$$ \be \bk=\bk^r(\bx,\bp):=\frac{\bx}{\vbx}, \quad
  \bB= \bB^r(\bx,\bp):=\frac{\bx^2\bp-(\bp\bx) \bx}{\vbx },\label{blm} \ee
  with the inverse map $\cR_i^r$ given by
$$
  \bx=\sqrt{2H}\bk,\quad \bp=-\sqrt{2H}T\bk+\frac{\bB}{\sqrt{2H}}.$$
  
The direct calculation of the bold-faced Poisson brackets gives 
   
    \be \an t^r,h^r\b=1,\quad
 \an h^r,\bk^r\b=\boldsymbol{0},\quad \an t^r,\bk^r\b=\boldsymbol{0},\quad 
 \an h^r,\bB^r\b=\boldsymbol{0},\quad \an t^r,\bB^r\b=\boldsymbol{0},\label{sepr}\ee \be\an B^r_i,B^r_j\b=B^r_ik^r_j-B^r_jk^r_i,\quad
\an k^r_i,B^r_j\b=\delta_{ij}-k^r_ik^r_j,\quad \an k^r_i,k^r_j\b=0, \quad i,j=1,...,n.\label{dirbr}\ee 
 
Comparing   \eqref{sep},\eqref{dirb} with \eqref{sepr},\eqref{dirbr}, we conclude that the map $\R^r$ is indeed the symplectomorphism.

\begin{rem} {\footnotesize 
In Section 3, we shall present as the main technical result of this paper an explicit  construction of four symplectomorphisms from $M_n$ to $P_n$ denoted respectively as $\cR^M,\cR^{yM},\cR^C,\cR^y$. Those four symplectomorhisms will have all geometrical interpretation. It is perhaps worth  pointing out that there exist also symplectomorphisms which do not have geometric interpretation, like, for example,  $\R^r$    where the Hamiltonian
$h^r(\bx,\bp)$ does not have a kinetic term. 
The  reason why we have introduced $\R^r$ is the fact  that  the $\bx,\bp$-depending vectors $\bk^r(\bx,\bp)$, $\bB^r(\bx,\bp)$ will play an important technical role throughout the paper. }
\end{rem}

\section{Explicit   canonical transformations}
  \subsection{Calogero-Moser system in the flat space} 
  In this  section, we show that the Calogero-Moser system in the flat space is symplectomorphic to the referential maximally superintegrable system of Section 2.
  
  \medskip

Spherically symmetric Calogero-Moser dynamical system $(M_n,\Om_n,h^{M})$ is defined by the Hamiltonian 
\be h^{M}(\bx,\bp):=\jp\bp^2+\jp\frac{\ga^2}{\bx^2}.\label{hcm}\ee
The flows generated by \eqref{hcm} are complete due to the conservation of energy and the fact, that the Calogero-Moser Hamiltonian is the sum of two positive terms, therefore neither kinetic nor potential energy may diverge within a given flow characterized by some conserved value of energy. This means that a particle can never reach  singularity at $\bx=\bo$
nor develop an unbounded velocity  which would be necessary in order to reach infinity in a finite time.

\medskip 

   Following \eqref{hamp}, the  Calogero-Moser  Hamiltonian $h^{M}(\bx,\bp)$ 
has the geometric interpretation as the Hamiltonian of the $0+1$-dimensional
$\sigma$-model. Indeed, it corresponds to the motion of a charged particle
in a flat space $\br^n$ and in a repulsive centrally symmetric electric potential $V(\bx)=\jp \ga^2 \bx^{-2}$.

\medskip

\begin{rem} {\footnotesize
The Calogero-Moser dynamical system is sometimes referred to as the conformal field theory in $0+1$-dimensions. The reason for this interpretation is the fact that the conformal group in $0+1$-dimension is $SL(2,\br)$ and it is infinitesimally generated via the Poisson brackets by the Hamiltonian  $h^{M}$, a dilation
charge $D=-\jp\bp\bx$ and a special conformal transformation charge $C=\jp \bx^2$.  It is easy to verify that the Poisson brackets of those generators form
the $sl(2,\br)$  Lie algebra
$$\an h^{M},D\b =h^{M},\quad \an C,D\b =-C,\quad \an h^{M},C\b = 2D.  $$
}
\end{rem}

\medskip

It is well-known that the flat  Calogero-Moser system is superintegrable in three dimensions \cite{OJ}, our  goal is now to show that the   $n$-dimensional version $(M_n,\Om_n,h^{M})$ is also superintegrable and, moreover, it is  symplectomorphic precisely  to the superintegrable  referential dynamical system\footnote{It should be noted    that  two given  spherically symmetric maximally superintegrable models need not be necessarily   symplectomorphic to each other.
In particular, the phase space of one of them   may   be symplectomorphic to our referential phase space $P_n$ but the
phase space of the other  may be rather symplectomorphic to a   $\mathbb Z$-quotient of  $P_n$ (in this case the symplectic half-plane $H,T$ becomes a symplectic half-cylinder with $T$ becoming an angle variable). Other scenarios  are also possible. 
}  $(P_n,\om_n,H)$. 
For that, consider a map $\cR^{M}:M_n\to P_n$ defined as
\be H=h^{M}(\bx,\bp)=\jp(\bp^2+\ga^2\bx^{-2} ),\quad   T=t^{M}(\bx,\bp):=\frac{\bp\bx}{\bp^2+\ga^2\bx^{-2}},\label{fom}\ee
\be  \label{fom2}
 \bk=\bk^{M}(\bx,\bp):= \bk^r(\bx)\cos{\Psi^{M}(\bx,\bp)}-\frac{\bB^r(\bx,\bp)} {\vert \bB^r(\bx,\bp)\vert} \sin{\Psi^{M}(\bx,\bp)},\ee \be  \label{fom3}
  \bB= \bB^{M}(\bx,\bp):=\bB^r(\bx,\bp) \cos{\Psi^{M}(\bx,\bp)}+{\vert \bB^r(\bx,\bp)\vert}\bk^r(\bx) \sin{\Psi^{M}(\bx,\bp)},  \ee
where
$$  \Psi^{M}(\bx,\bp)= \frac{\vert \bB^r(\bx,\bp)\vert}{\sqrt{\vert \bB^r(\bx,\bp)\vert^2+\ga^2}}\arctan{\frac{\bp\bx}{\sqrt{\vert \bB^r(\bx,\bp)\vert^2+\ga^2}}}.
$$
We verify easily that it holds
$$\left(\bk^{M}(\bx,\bp)\right)^2=1,\quad \bk^{M}(\bx,\bp)\bB^{M}(\bx,\bp)=0,$$
we thus observe that the map $\cR^M$ is indeed  from $M_n$ to $P_n$. Moreover, the map $\cR^M$ is evidently defined on the whole $M_n$ and is  smooth everywhere on $M_n$.

\medskip

Now consider a map
$\cR_i^M:P_n\to M_n$ defined by
\begin{subequations}
 \label{cmn}
 \begin{align}
  \bx&=
 N^{M}(H,T,\vbB)\left(\bk\cos{\Psi^{M}(H,T,\vbB) }+\frac{\bB}{\vbB}\sin{\Psi^{M}(H,T,\vbB) }\right)\\
  \bp&= \frac{(2HT\bk+\bB)\cos{\Psi^{M}(H,T,\vbB) }+(2HT\frac{\bB}{\vbB}-\vbB\bk)\sin{\Psi^{M}(H,T,\vbB) }}{ N^{M}(H,T,\vbB)} ,\end{align}
\end{subequations}
where
$$ N^{M}(H,T,\vbB)= \frac{\sqrt{4H^2T^2+\bB^2+\ga^2}}{\sqrt{2H}},$$
$$ \Psi^{M}(H,T,\vbB) =\frac{\vert \bB\vert}{\sqrt{\bB^2+\ga^2}}\arctan{\frac{2HT}{\sqrt{\bB^2+\ga^2}}}.$$

\medskip

The map $\cR^M_i$ is evidently well defined on the whole $P_n$ and it is everywhere smooth because the apparent singularity 
at     $\vert \bB\vert=0$ is smoothly removable due to the multiplication by $\sin{\Psi^{M}}$.

\medskip 

We readily verify that
$$ \cR^M\circ \cR^M_i={\rm Id}_{P_n}, \quad \cR^M_i\circ \cR^M={\rm Id}_{M_n},$$
which means that the both maps $\cR^M,\cR^M_i$ are diffeomorphisms inverse to each other. 

\medskip 

A direct calculation of the bold-faced Poisson brackets then gives 
    \be \an t^{M},h^{M}\b=1,\quad 
 \an h^{M},\bk^{M}\b=  \an t^{M},\bk^{M}\b= 
 \an h^{M},\bB^{M}\b= \an t^{M},\bB^{M}\b=\boldsymbol{0},\label{sepcm}\ee \be\an B^{M}_i,B^{M}_j\b=B^{M}_ik^{M}_j-B^{M}_jk^{M}_i,\ 
\an k^{M}_i,B^{M}_j\b=\delta_{ij}-k^{M}_ik^{M}_j,\  \an k^{M}_i,k^{M}_j\b=0.\label{cmbr}\ee 
Comparing   \eqref{sepcm},\eqref{cmbr} with \eqref{sep},\eqref{dirb}, we conclude that the diffeomorphism $\R^{M}$ is in fact the symplectomorphism.
Said in other words,
we have just shown that
the flat Calogero-Moser system $(M_n,\Om_n,h^{M})$ is symplectomorphic to the referential dynamical system $(P_n,\om_n,H)$ via the symplectomorphism $\cR^{M}$, in particular, we have  
$$ H=h^{M}(\bx,\bp)=\jp(\bp^2+\ga^2\bx^{-2} ).$$

 \medskip

If  we   interpret the variable $T$ in \eqref{cmn} as time and $H,\bk,\bB$ as constant quantities, the inverse symplectomorphism \eqref{cmn}
can be checked to be  the solution of the Calogero-Moser Hamiltonian equations of motion
 \be\dot \bx =\an\bx,h^{M}\b=\bp,\quad \dot \bp=\an\bp,h^{M}\b=\frac{\ga^2\bx}{(\bx^{2})^2}.\label{cmem}\ee
 
 In reality, we have used this very fact   to find the explicit form \eqref{fom}, \eqref{fom2} and \eqref{fom3} of the
 symplectomorphism $\cR^M$.  We  have first found the general solution \eqref{cmn} of the Calogero-Moser Hamiltonian equations of motions \eqref{cmem}, we interpreted the time as the variable canonically conjugated to the Hamiltonian and then we expressed $H,T,\bk,\bB$ as the functions of $\bp,\bx$. It was of course not clear from the outset what kind of the bold-faced Poisson brackets would obey those functions,   but it turned out eventually that they do obey those of the referential dynamical system $(P_n,\om_n,H)$. It is this circumstance  which makes the spherically symmetric Calogero-Moser model   propitious to admit point particle T-duals.
   \subsection{Calogero-Moser system in the hyperbolic space} 
     In this  section, we show that the Calogero-Moser system in the   space of constant negative curvature is symplectomorphic to the referential maximally superintegrable system of Section 2.

 \medskip

   Equip the space $\br^n$ with a metric
   \be g_{jk}=\delta_{jk}-\frac{\al^2x^jx^k}{1+\al^2\bx^2}.\label{ag}\ee
  The
  scalar curvature of the metric \eqref{ag} is constant 
$$ R=-n(n-1)\al^2;$$
  the space $\br^n$ equipped with the metric \eqref{ag} is then called the hyperbolic space  or the space of negative constant curvature.
  
  \medskip 
  
  Note that the inverse metric tensor reads
 $$ g^{jk}(\bx)=\delta^{jk}+\al^2x^jx^k, $$
  therefore the Hamiltonian \eqref{hamp}   of a charged point particle moving
  in the background \eqref{ag} and feeling the electric potential
  $V(\bx)=\jp\frac{\ga^2}{\bx^2}$
  is 
   \be h^{yM}(\bx,\bp)= \jp g^{jk}(\bx)p_jp_k+V(\bx)= \jp\left(\bp^2+\frac{\ga^2}{\bx^2}+\al^2(\bp\bx)^2\right).\label{a}\ee
   We thus observe, that the dynamical system $(M_n,\Om_n,h^{yM})$ is an $\al$-deformation of the flat Calogero-Moser dynamics described in the previous section, the deformation which physically corresponds to switching on  the negative  constant curvature.
   
   \medskip 
   
  Note that the flows generated by  \eqref{a} are again complete due to
   a variant of the argument given in the previous section for the case $\al=0$. Indeed, the hyperbolic Calogero-Moser Hamiltonian is the sum of   positive terms, therefore neither kinetic nor potential energy may diverge within a given flow characterized by some conserved value of energy.

\medskip
 
Our goal is to show that the hyperbolic  model $(M_n,\Om_n,h^{yM})$ is symplectomorphic to the referential dynamical system $(P_n,\om_n,H)$. 
For that, consider a map $\cR^{yM}:M_n\to P_n$ defined as
\be H=h^{yM}(\bx,\bp)=\jp(\bp^2+\ga^2\bx^{-2}+\al^2(\bp\bx)^2),\label{hham}\ee 
\be T=t^{yM}(\bp,\bx)=\frac{{\rm argtanh}\frac{\al\bp\bx}{\sqrt{\bp^2+\ga^2\bx^{-2}+\al^2(\bp\bx)^2}}}{\al\sqrt{\bp^2+\ga^2\bx^{-2}+\al^2(\bp\bx)^2}},\label{htoa}\ee
 \be \label{fomh2}
 \bk=\bk^{yM}(\bx,\bp):= \bk^r(\bx)\cos{\Psi^{yM}(\bx,\bp)}-\frac{\bB^r(\bx,\bp)} {\vert \bB^r(\bx,\bp)\vert} \sin{\Psi^{yM}(\bx,\bp)},\ee \be  \label{fomh3}
  \bB= \bB^{yM}(\bx,\bp):=\bB^r(\bx,\bp) \cos{\Psi^{yM}(\bx,\bp)}+{\vert \bB^r(\bx,\bp)\vert}\bk^r(\bx) \sin{\Psi^{yM}(\bx,\bp)},  \ee
where
$$ \Psi^{yM}(\bx,\bp)= \frac{\vert \bB^r(\bx,\bp)\vert}{\sqrt{\vert \bB^r(\bx,\bp)\vert^2+\ga^2}}\arctan{\frac{\bp\bx}{\sqrt{\vert \bB^r(\bx,\bp)\vert^2+\ga^2}}}.$$
We verify easily that it holds
$$ \left(\bk^{yM}(\bx,\bp)\right)^2=1,\quad \bk^{yM}(\bx,\bp)\bB^{yM}(\bx,\bp)=0,$$ 
we thus observe that the map $\cR^{yM}$ is indeed  from $M_n$ to $P_n$. Moreover, the map $\cR^{yM}$ is evidently defined on the whole $M_n$ and is   smooth everywhere on $M_n$.

\medskip

Now consider a map
$\cR_i^{yM}:P_n\to M_n$ defined by
\be 
  \bx=
 N^{yM}(H,T,\vbB)\left(\bk\cos{\Psi^{yM}(H,T,\vbB) }+\frac{\bB}{\vbB}\sin{\Psi^{yM}(H,T,\vbB) }\right)\\
  \label{cmhna}\ee
$$  \bp= \frac{\left(\sqrt{2H}\tanh{(\al\sqrt{2H}T})\bk+\al\bB\right)\cos{\Psi^{yM}(H,T,\vbB) }}{\al N^{yM}(H,T,\vbB)} +$$
\be+\frac{\left(\sqrt{2H}\tanh{(\al\sqrt{2H}T)}\frac{\bB}{\vbB}-\al\vbB\bk\right)\sin{\Psi^{yM}(H,T,\vbB)}}{\al N^{yM}(H,T,\vbB)},\label{cmhnb}  
\ee
where
$$ N^{yM}(H,T,\vbB) =\sqrt{\frac{(\bB^2+\ga^2)\cosh^2{(\al\sqrt{2H}T)}}{2H}+\frac{\sinh^2{(\al\sqrt{2H}T)}}
{\al^2}}
,$$
and
$$\Psi^{yM}(H,T,\vbB) =\frac{\vert \bB\vert}{\sqrt{\bB^2+\ga^2}}\arctan{\frac{\sqrt{2H}\tanh{(\al\sqrt{2H}T)}}{\al\sqrt{\bB^2+\ga^2}}}.$$

\medskip

The map $\cR^{yM}_i$ is evidently well defined on the whole $P_n$ and it is everywhere smooth because the apparent singularity 
at     $\vert \bB\vert=0$ is smoothly removable due to the multiplication by $\sin{\Psi^{yM}}$.

\medskip 

We readily verify that
$$ \cR^{yM}\circ \cR^{yM}_i={\rm Id}_{P_n}, \quad \cR^{yM}_i\circ \cR^{yM}={\rm Id}_{M_n},$$
which means that the both maps $\cR^{yM},\cR^{yM}_i$ are diffeomorphisms inverse to each other. 

\medskip

A direct   calculation of the bold-faced Poisson brackets then gives 
    \be \an t^{yM},h^{yM}\b=1,\quad 
 \an h^{yM},\bk^{yM}\b=  \an t^{yM},\bk^{yM}\b= 
 \an h^{yM},\bB^{yM}\b= \an t^{yM},\bB^{yM}\b=\boldsymbol{0},\label{sepcmh}\ee \be\an B^{yM}_i,B^{yM}_j\b=B^{yM}_ik^{yM}_j-B^{yM}_jk^{yM}_i,\ 
\an k^{yM}_i,B^{yM}_j\b=\delta_{ij}-k^{yM}_ik^{yM}_j,\  \an k^{yM}_i,k^{yM}_j\b=0.\label{cmhb}\ee 
Comparing   \eqref{sepcmh},\eqref{cmhb} with \eqref{sep},\eqref{dirb},  we conclude that the diffeomorphism $\R^{yM}$ is in fact the symplectomorphism.
Said in other words, 
we have just shown that
the hyperbolic Calogero-Moser system $(M_n,\Om_n,h^{yM})$ is symplectomorphic to the referential dynamical system $(P_n,\om_n,H)$ via the symplectomorphism $\cR^{yM}$, in particular, we have  
$$ H=h^{yM}(\bx,\bp)=\jp\left(\bp^2+\ga^2\bx^{-2}+\al^2(\bp\bx)^2 \right).$$

 \medskip

If  we   interpret the variable $T$ in \eqref{cmhna} and \eqref{cmhnb} as time and $H,\bk,\bB$ as constant quantities, the inverse symplectomorphism \eqref{cmhna} and \eqref{cmhnb}
can be checked to be  the solution of the hyperbolic Calogero-Moser Hamiltonian equations of motion
\be \dot\bx=\an\bx,h^{yM}\b=\bp+\al^2(\bp\bx)\bx,\quad \dot \bp=\an\bp,h^{yM}\b=\frac{\ga^2\bx}{(\bx^2)^2}-\al^2(\bp\bx)\bp.\label{cmhem}\ee



 In reality, we have used this very fact   to find the explicit form \eqref{hham},\eqref{htoa},\eqref{fomh2} and \eqref{fomh3} of the
 symplectomorphism $\cR^{yM}$.  We  have first found the general solution \eqref{cmhna}, \eqref{cmhnb} of the Calogero-Moser Hamiltonian equations of motions \eqref{cmhem}, we interpreted the time as the variable canonically conjugated to the Hamiltonian and then we expressed $H,T,\bk,\bB$ as the functions of $\bp,\bx$. It was of course not clear from the outset what kind of the bold-faced Poisson brackets would obey those functions,  but it turned out eventually that they do obey those of the referential dynamical system $(P_n,\om_n,H)$. It is this circumstance  which makes the hyperbolic spherically symmetric Calogero-Moser model   propitious to admit point particle T-duals.

 \subsection{Repulsive Coulomb potential in the flat space}
    In this  section, we show that the standard repulsive Coulomb  system in the flat space is symplectomorphic to the referential maximally superintegrable system of Section 2.
    
    \medskip

  We   consider  a Hamiltonian
  \be h^C(\bx,\bp):=\jp \bp^2  +\frac{\bt^2} {\vbx},\label{cou}\ee 
  which has a natural physical interpretation  in the dimension $n=3$   because $\frac{\bt^2}{\vbx}$ is
 the repulsive Coulomb potential in the flat three-dimensional space.
 
  \medskip 
  
The flows generated by \eqref{cou} can be shown to be complete by essentially the same argument as in the Calogero-Moser case.    

 \medskip
 
  Our goal is to show that the flat Coulomb system $(M_n,\Om_n,h^{C})$ is symplectomorphic to the referential dynamical system $(P_n,\om_n,H)$.
For that, we consider a map
  $\cR^{C}:M_n\to P_n$   given by 
 \be H=h^C(\bx,\bp)=\jp \bp^2  +\frac{\bt^2} {\vbx},\label{hca}\quad  T=t^C(\bx,\bp):=\tau^C(\bp\bx,h^C(\bx,\bp), K^C 
(\bx,\bp)), \ee
\be \label{foc2}
\bk=\bk^{C}(\bx,\bp):= \frac{(\bt^2\vbx+\vert \bB^r(\bx,\bp)\vert^2)\bk^r(\bx)-(\bp\bx)  \bB^r(\bx,\bp)}{K^C(\bx,\bp)\vbx}\ee
\be \label{foc3}
\bB=\bB^{C}(\bx,\bp):= \frac{(\bt^2\vbx+\vert \bB^r(\bx,\bp)\vert^2)\bB^r(\bx)+\vert \bB^r(\bx,\bp)\vert^2(\bp\bx)  \bk^r(\bx)}{K^C(\bx,\bp)\vbx}\ee
 where $\bk^r(\bx)$,   $\bB^r(\bx,\bp)$ were defined in \eqref{blm} and
 $$ K^C(\bx,\bp):=\sqrt{2h^C(\bx,\bp)\vert \bB^r(\bx,\bp)\vert^2+\bt^4},$$
 \be \tau^C(\bp\bx,H,K):=\frac{\bp\bx}{2H}+\frac{\bt^2}{\sqrt{2H}^3}{\rm argsinh}\left(\frac{\sqrt{2H}}{K}\bp\bx\right).\label{tauc}\ee
 
  We verify easily that it holds
$$ \left(\bk^{C}(\bx,\bp)\right)^2=1,\quad \bk^{C}(\bx,\bp)\bB^{C}(\bx,\bp)=0,$$
we thus observe that the map $\cR^C$ is indeed  from $M_n$ to $P_n$. Moreover, the map $\cR^C$ is evidently defined on the whole $M_n$ and it is   smooth everywhere on $M_n$.

\medskip

Now consider a map $\cR_i^{C}:P_n\to M_n$ defined by
\be \bx=\frac{(K^2+\bt^2\sqrt{K^2+2HJ^2(T,H,K)})\bk+2HJ(T,H,K)\bB}{2HK},\label{ica} \ee \be \bp=\frac{2H}{K}\left(\bB+\frac{\bt^2(J(T,H,K)\bk-\bB)}{\bt^2+\sqrt{K^2+2HJ^2(T,H,K)}}\right),\label{icb}\ee
where
\be K\equiv \sqrt{2H\vbB^2+\bt^4}, \quad \tau^C(J(T,H,K),H,K)=T.\label{kjt}\ee
 Note that the second equation of \eqref{kjt} is the definition of the function $J(T,H,K)$, that is $J(T,H,K)$ is the function inverse to $\tau^C(\bp\bx,H,K)$ viewed as the function of the first argument. The fact that this   inverse function
 exists follows from 
    taking 
a partial derivative of $\tau^C$ with respect to $\bp\bx$ for fixed $H>0$ and $K\geq\bt^2$. Indeed, we find from \eqref{tauc}
$$ \frac{\d\tau^C(\bp\bx,H,K)}{\d(\bp\bx)}=\frac{1}{2H}+\frac{\bt^2}{2H\sqrt{K^2+2H(\bp\bx)^2}}>0.$$
  Therefore, for $H,K$ fixed, the  function $\tau^C(\bp\bx,H,K)$ is increasing as the function of $\bp\bx$ and it admits the smooth inverse function $J(T,H,K)$.  
  
  \medskip
  
  The map $\cR^{C}_i$ is evidently well defined on the whole $P_n$ and it is everywhere smooth.    Moreover,
we readily verify that
$$ \cR^{C}\circ \cR^{C}_i={\rm Id}_{P_n}, \quad \cR^{C}_i\circ \cR^{C}={\rm Id}_{M_n},$$
which means that the both maps $\cR^{C},\cR^{C}_i$ are diffeomorphisms inverse to each other.



 \medskip
 



 A direct   calculation of the bold-faced Poisson brackets finally gives 
    \be \an t^{C},h^{C}\b=1,\quad 
 \an h^{C},\bk^{C}\b=  \an t^{C},\bk^{C}\b= 
 \an h^{C},\bB^{C}\b= \an t^{C},\bB^{C}\b=\boldsymbol{0},\label{cea}\ee \be\an B^{C}_i,B^{C}_j\b=B^{C}_ik^{C}_j-B^{C}_jk^{C}_i,\ 
\an k^{C}_i,B^{C}_j\b=\delta_{ij}-k^{C}_ik^{C}_j,\  \an k^{C}_i,k^{C}_j\b=0.\label{ceb}\ee 
\medskip

Comparing   \eqref{cea},\eqref{ceb} with \eqref{sep},\eqref{dirb}, we conclude that  the diffeomorphism $\R^{C}$ is in fact the symplectomorphism.
Said in other words, 
we have just shown that
the flat   Coulomb system $(M_n,\Om_n,h^{C})$ is symplectomorphic to the referential dynamical system $(P_n,\om_n,H)$ via the symplectomorphism $\cR^{C}$, in particular, we have  
$$ H=h^{C}(\bx,\bp)=\jp \bp^2+\frac{\bt^2}{\vbx}  .$$

\medskip

If in \eqref{ica}, \eqref{icb}  we   interpret the variable $T$ as time and $H,\bk,\bB$ as constant quantities, the inverse symplectomorphism \eqref{ica}, \eqref{icb}
can be checked to be  the solution of the flat Coulomb  equations of motion
 \be 
   \dot\bx=\an\bx,h^C\b=\bp,\quad 
  \label{bemb}\dot \bp=\an\bp,h^C\b= \frac{\bt^2\bx}{\vbx^3\ }   .
\ee
 
 
In reality, we have used this very fact   to find the explicit form \eqref{hca},\eqref{foc2},\eqref{foc3} of the
 symplectomorphism $\cR^C$.  We  have first found the general solution \eqref{ica}, \eqref{icb} of the flat Coulomb Hamiltonian equations of motions \eqref{bemb}, we interpreted the time as the variable canonically conjugated to the Hamiltonian and then we expressed $H,T,\bk,\bB$ as the functions of $\bp,\bx$. It was of course not clear from the outset what kind of the bold-faced Poisson brackets would obey those functions,   but it turned out eventually that they do obey those of the referential dynamical system $(P_n,\om_n,H)$. It is this circumstance  which makes the flat Coulomb model   propitious to admit point particle T-duals.
 
 \medskip 
 
 \begin{rem}{\footnotesize The reader might not have recognized
 in \eqref{ica} and \eqref{icb} the standard solution of the Coulomb (or Kepler) problem as we have intentionally avoided to employ the spherical coordinates. Indeed,
  any use of local coordinate systems like the spherical coordinates would obscure our task   to find the global symplectomorphism relating the Coulomb model to the referential one.}
  
  \end{rem}
 \subsection{ Repulsive Coulomb potential in the hyperbolic space}
   In this  section, we show that the repulsive Coulomb model in the   space of constant negative curvature is symplectomorphic to the referential maximally superintegrable system of Section 2.
 
 \medskip

   We consider  a Hamiltonian   of a charged point particle moving
  in the hyperbolic space  background \eqref{ag} and feeling the electric potential
  $V(\bx)=\frac{\bt^2\sqrt{1+\al^2\bx^2}}{\vbx}$:
 \be h^{y}(\bx,\bp)=\jp\left(\bp^2 +\al^2(\bp\bx)^2\right)+\frac{\bt^2\sqrt{1+\al^2\bx^2}}{\vbx}-\al\bt^2, \quad \al>0.\label{hrch}\ee
   The flows generated by  \eqref{hrch} are  complete due to
 essentially the same argument as in the previous sections.

   \medskip
 
Note that the Hamiltonian $h^{y}$ is well defined in any number of dimensions, but in the dimension $n=3$ it has a natural physical interpretation as
 the repulsive Coulomb potential in the space of negative constant curvature. Indeed, in three dimensions the Laplace-Beltrami operator in the background \eqref{ag} acts on the potential
$V(\bx)=\frac{\bt^2\sqrt{1+\al^2\bx^2}}{\vbx}-\al\bt^2$ with the result
 $$ \Delta_{LB}\frac{ \sqrt{1+\al^2\bx^2}}{\vbx}\equiv\frac{1}{\sqrt{\det{g}}}\d_{x_j}\left(\sqrt{\det{g}}g^{jk}\d_{x_j}\frac{ \sqrt{1+\al^2\bx^2}}{\vbx}\right)=-4\pi\delta(\bx).$$
 
 \medskip
 
 It is well-known that the 
   hyperbolic Coulomb system is superintegrable in three dimensions \cite{E90,BM05,BEHR09,Q16,BBH19},
   our goal is to show a little bit more than this, that is to show that its   $n$-dimensional version 
   $(M_n,\Om_n,h^{y})$ is symplectomorphic   to the referential dynamical system $(P_n,\om_n,H)$.
For that, we consider a map
  $\cR^{y}:M_n\to P_n$   given by 
   
 \be H=h^{y}(\bp,\bx)=\jp\left(\bp^2 +\al^2(\bp\bx)^2\right)+\frac{\bt^2\sqrt{1+\al^2\bx^2}}{\vbx}-\al\bt^2,\label{hy}\ee\be T=t^y(\bx,\bp):=\tau^y(\bp\bx,h^y(\bx,\bp), \vert \bB^r(\bx,\bp)\vert), \label{ty}\ee
\be \label{foy2}
\bk=\bk^{y}(\bx,\bp):=\frac{\bt^2\bk^r(\bx,\bp)}{K^y(\bx,\bp)}+\sqrt{1+\al^2\bx^2} \frac{ \vert \bB^r(\bx,\bp)\vert^2\bk^r(\bx)-(\bp\bx)  \bB^r(\bx,\bp)}{K^y(\bx,\bp)\vbx},\ee
\be \label{foy3}
\bB=\bB^{y}(\bx,\bp):=\frac{\bt^2\bB^r(\bx,\bp)}{K^y(\bx,\bp)} +\vert \bB^r(\bx,\bp)\vert^2\sqrt{1+\al^2\bx^2}\frac{ \bB^r(\bx,\bp)+ (\bp\bx)  \bk^r(\bx)}{K^y(\bx,\bp)\vbx}\ee
 where $\bk^r(\bx)$,   $\bB^r(\bx,\bp)$ were defined in \eqref{blm} and
$$  K^y(\bx,\bp):=\sqrt{2h^y(\bx,\bp)\vert\bB^r(\bx,\bp)\vert^2+(\bt^2+\al\vert\bB^r(\bx,\bp)\vert^2)^2},$$
$$ \tau^{y}(\bp\bx,H,B) 
 :=$$\be=\frac{{\rm Argtanh} \left(\frac{A^\al(B,K)(K-\bt^2)\bp\bx}{BK+\sqrt{B^2K^2+(\bp\bx)^2(K^2-\bt^4)}}\right)}{\al\sqrt{2H}}-\frac{{\rm Argtanh} \left(\frac{A^{-\al}(B,K)(K-\bt^2)\bp\bx}{BK+\sqrt{B^2K^2+(\bp\bx)^2(K^2-\bt^4)}}\right)}{\al\sqrt{2H+4\al\bt^2}}\label{tyhb}\ee 
  \be A^{\pm\al}(B,K)=\sqrt{\frac{K+(\bt^2\pm \al B^2)}{
  K -(\bt^2\pm \al B^2) }}.\label{apm}\ee
  
  \medskip
  We verify  that it holds
$$ \left(\bk^{y}(\bx,\bp)\right)^2=1,\quad \bk^{y}(\bx,\bp)\bB^{y}(\bx,\bp)=0,$$
we thus observe that the map $\cR^y$ is indeed  from $M_n$ to $P_n$. Moreover, the map $\cR^y$ is  defined on the whole $M_n$ (see Appendix) and it is   smooth everywhere on $M_n$.

\medskip

Now consider a map $\cR_i^{y}:P_n\to M_n$ defined by
  \be
     \label{iha}
  \bx=\frac{\vbB^2\bk\cos{\Psi^y(T)}+\vbB\bB \sin{\Psi^y(T)}}{\sqrt{(K\cos{\Psi^y(T)}-\bt^2)^2-\al^2\vbB^4}},\ee   $$ \bp=\frac{\sqrt{(K\cos{\Psi^y(T)}-\bt^2)^2-\al^2\vbB^4}}{\vbB^2}\times $$\be\times\left((\bB+Y(T)\bk)\cos{\Psi^y(T)}+(Y(T)\frac{\bB}{\vbB}-\vbB \bk)\sin{\Psi^y(T)}\right),\label{ihb}\ee
where  
$$ K=\sqrt{2H\vbB^2+(\al\vbB^2+\bt^2)^2},$$
$$ \sin{\Psi^y(T)}:=\frac{ (K-\bt^4K^{-1})Y(T,H,\vbB)}{ \sqrt{K^2\vbB^2+(K^2-\bt^4)Y^2(T,H,\vbB) }+ \vbB\bt^2},$$
$$ \cos{\Psi^y(T)}:=\frac{\bt^2}{K}+  \frac{(K-\bt^4K^{-1})  \vbB }{\sqrt{K^2\vbB^2+(K^2-\bt^4)Y^2(T,H,\vbB)}+  \vbB\bt^2}.$$
Here the function $Y(T,H,B)$ is inverse to $\tau^y(\bp\bx,H,B)$ viewed as the function of the first argument (with $H,B$ fixed). The proof that this   inverse function
 exists is presented in the Appendix.   
 
  \medskip
  
  The map $\cR^{y}_i$ is   well defined on the whole $P_n$ and it is everywhere smooth.    Moreover,
we readily verify that
$$ \cR^{y}\circ \cR^{y}_i={\rm Id}_{P_n}, \quad \cR^{y}_i\circ \cR^{y}={\rm Id}_{M_n},$$
which means that the both maps $\cR^{y},\cR^{y}_i$ are diffeomorphisms inverse to each other. 
    
 \medskip
   
 A direct (and tedious) calculation of the bold-faced Poisson brackets finally gives 
    \be \an t^{y},h^{y}\b=1,\quad 
 \an h^{y},\bk^{y}\b=  \an t^{y},\bk^{y}\b= 
 \an h^{y},\bB^{y}\b= \an t^{y},\bB^{y}\b=\boldsymbol{0},\label{ceha}\ee \be\an B^{y}_i,B^{y}_j\b=B^{y}_ik^{y}_j-B^{y}_jk^{y}_i,\ 
\an k^{y}_i,B^{y}_j\b=\delta_{ij}-k^{y}_ik^{y}_j,\  \an k^{y}_i,k^{y}_j\b=0.\label{cehb}\ee 
\medskip

Comparing   \eqref{ceha},\eqref{cehb} with \eqref{sep},\eqref{dirb}, we conclude that  the diffeomorphism $\R^{y}$ is in fact the symplectomorphism.
Said in other words, 
we have just shown that
the hyperbolic Coulomb system $(M_n,\Om_n,h^{y})$ is symplectomorphic to the referential dynamical system $(P_n,\om_n,H)$ via the symplectomorphism $\cR^{y}$, in particular, we have  
$$ H= h^{y}(\bx,\bp)=\jp\left(\bp^2 +\al^2(\bp\bx)^2\right)+\frac{\bt^2\sqrt{1+\al^2\bx^2}}{\vbx}-\al\bt^2. $$

\medskip

If  we   interpret the variable $T$ in \eqref{iha} and \eqref{ihb} as time and $H,\bk,\bB$ as constant quantities, the inverse symplectomorphism  \eqref{iha} and \eqref{ihb}
can be checked to be  the solution of the hyperbolic Coulomb Hamiltonian equations of motion
  \begin{subequations}
 \label{beh}
 \begin{align}
     \label{beha}
   \dot\bx&=\an\bx,h^y\b=\bp+\al^2(\bp\bx)\bx,\\
  \label{behb}\dot \bp&=\an\bp,h^y\b= \frac{\bt^2\bx}{\vbx^3\sqrt{1+\al^2\bx^2}}  -\al^2(\bp\bx)\bp.\end{align}
\end{subequations}

In reality, we have used this very fact   to find the explicit form \eqref{hy},\eqref{ty}, \eqref{foy2},\eqref{foy3} of the
 symplectomorphism $\cR^y$.  We  have first found the general solution \eqref{iha}, \eqref{ihb}  of the hyperbolic Coulomb Hamiltonian equations of motions \eqref{beh}, we interpreted the time as the variable canonically conjugated to the Hamiltonian and then we expressed $H,T,\bk,\bB$ as the functions of $\bp,\bx$. It was of course not clear from the outset what kind of the bold-faced Poisson brackets would obey those functions,   but it turned out eventually that they do obey those of the referential dynamical system $(P_n,\om_n,H)$. It is this circumstance  which makes the hyperbolic
 Coulomb model   propitious to admit point particle T-duals.
  
 

  \section{Discussion, conclusions and outlook}

  In the preceding  Section 3, we have shown that four superintegrable dynamical systems, i.e. flat and hyperbolic Calogero-Moser and flat and hyperbolic Coulomb, are all symplectomorphic to the referential dynamical system 
  $(P_n,\Om_n,H)$ introduced in Section 2. We have constructed explicitely the respective symplectomorphisms
  $\cR^M,\cR^{yM},\cR^{C},\cR^y: M_n\to P_n$ as well as the inverse symplectomorphisms   $\cR_i^M,\cR_i^{yM},\cR_i^{C},\cR_i^y: P_n\to M_n$.
  The T-duality symplectomorphisms relating those four models  are given by the compositions of one original and one inverse symplectomorphism. 
   For example, the  hyperbolic Calogero-Moser is related to the flat Coulomb  by the composed  T-duality 
  symplectomorphism $\cR_i^{yM}\circ\cR^C$. The explicit formulas for those composed T-duality symplectomorphisms can be worked out straightforwardly but we do not list them because they are  cumbersome and, anyway,  not very illuminating.

  \medskip

Do some additional  $0+1$-dimensional $\si$-models
belong to the T-duality equivalence class consisting of the four dynamical systems that we have studied in detail? Very probably yes due to the result of Fradkin \cite{F67} which pioneered  the quest for superintegrability for   general spherically symmetric potentials. However, showing that a given spherically symmetric model is superintegrable is not sufficient, because, as it was  already remarked in Footnote 1,   not all  spherically symmetric  superintegrable models must be necessarily   symplectomorphic to our referential model $(P_n,\Om_n,H)$.
In particular, the phase spaces of some  spherically symmetric  superintegrable models may not be symplectomorphic to our referential phase space $P_n$ but they  may be rather symplectomorphic to a   $\mathbb Z$-quotient of  $P_n$ (in this case the symplectic half-plane $H,T$ becomes a symplectic half-cylinder with $T$ becoming an angle variable). Other scenarios are also possible and we expect that
new T-duality equivalence classes   can be discovered by following the track of superintegrability, including   cases where we abandon the requirement of the spherical symmetry.

\medskip

A big open issue is a  quantum status of the point particle T-duality. Although the  problem
looks  much easier than in the string case, still the
 high degree of nonlinearity of the  explicit symplectomorphisms obtained in Section 3 suggests, that it will not be an efortless task to settle it.
 
 \medskip


   \section*{Appendix}
   First we show that the  function \eqref{ty}
   is defined everywhere on 
   $M_n$. For that we rewrite it as
 $$ \tau^y(\bx,\bp)=  \frac{{\rm Argtanh}  \left(\frac{\tan{\frac{\Psi^y(\bx,\bp)}{2}}}{ \tan{\frac{\Phi^+(\bx,\bp) }{2}}}\right)}{\al\sqrt{2h^y(\bx,\bp)}}
- \frac{{\rm Argtanh}  \left(\frac{\tan{\frac{\Psi^y(\bx,\bp)}{2}}}{ \tan{\frac{\Phi^-(\bx,\bp) }{2}}}\right)}{\al\sqrt{2h^y(\bx,\bp)+4\al\bt^2}},$$
 where   \be \Psi^y(\bx,\bp) =  \arctan{\frac{\sqrt{1+\al^2\bx^2} \vert \bB^r(\bx,\bp)\vert(\bp\bx)}{\sqrt{1+\al^2\bx^2}\vert\bB^r(\bx,\bp)\vert^2+\bt^2\vbx} } ,\label{psch}\ee 
  \be \tan{\frac{\Phi^\pm(\bx,\bp)}{2}}=\sqrt{\frac{
  K^y(\bx,\bp )-(\bt^2\pm \al\vert\bB^r(\bx,\bp) \vert^2) }{K^y(\bx,\bp)+(\bt^2\pm \al\vert\bB^r(\bx,\bp) \vert^2)}}.\label{phh}\ee

 We show without difficulties that $0<\Phi^+(\bx,\bp)<\Phi^-(\bx,\bp)$ for all $(\bx,\bp)\in M_n$,
  therefore the domain of definition of $t^y(\bx,\bp)$  is given by all $(\bx,\bp)\in M_n$ which satisfy
 \be \left\vert  \tan{\frac{\Psi^y(\bx,\bp)}{2}}\right\vert<\tan{\frac{\Phi^+(\bx,\bp)}{2}} (<1).\label{tt}\ee
 Following \eqref{psch}, the image of the map $\Psi^y$ belongs  to the interval $]-\jp\pi,\jp\pi[$, whatever $(\bx,\bp)\in M_n$ we consider. We find also
  \be\cos{\Psi^y(\bx,\bp)}>\cos{\Phi^+(\bx,\bp) },\quad \forall (\bx,\bp)\in M_n .\label{cc}\ee
Indeed, it follows from \eqref{psch} and \eqref{phh}   
$$ \cos{\Psi^y(\bx,\bp)}=\frac{\vert \bB^r(\bx,\bp)\vert^2\sqrt{1+\al^2\bx^2}+\bt^2\vbx}{K^y(\bx,\bp)\vbx}, \  \cos{\Phi^+(\bx,\bp)}=\frac{\al\vert \bB^r(\bx,\bp)\vert^2\ +\bt^2}{K^y(\bx,\bp)},  $$
therefore  
$$K^y(\bx,\bp )\cos{\Psi^y(\bx,\bp)}-\bt^2-\al \vert \bB^r(\bx,\bp)\vert^2=\vert \bB^r(\bx,\bp)\vert^2\left(\frac{\sqrt{1+\al^2\bx^2}}{\vbx}-\al\right)  >$$\be >0= 
K^y(\bx,\bp) \cos{\Phi^+(\bx,\bp)}-\bt^2-\al \vert \bB^r(\bx,\bp)\vert^2.\label{ner} \ee
Finally, the inequality \eqref{ner} 
  implies \eqref{cc}, which in turn implies that the    relation   \eqref{tt} holds for all $(\bx,\bp)\in M_n$.

\vskip2pc

We  wish also  to show that, for
  $H,B$ fixed, the function 
  $t^y(\bp\bx,H,B)$ is  invertible as the function  of $(\bp\bx)$, which means that the partial derivative $\frac{\d t^y}{\d (\bp\bx)}$ must be positive.
  Set
  $$ z=\frac{(K-\bt^2)\bp\bx}{BK+\sqrt{B^2K^2+(\bp\bx)^2(K^2-\bt^4)}}$$
  and find that it holds
  $$\frac{\d z}{\d(\bp\bx)}= \frac{(K-\bt^2)BK}{(BK+\sqrt{B^2K^2+(\bp\bx)^2(K^2-\bt^4)})\sqrt{B^2K^2+(\bp\bx)^2(K^2-\bt^4)}}>0.$$
 Looking at \eqref{tyhb}, we therefore see that
 we have just to show
 \be \frac{\d \tau^y}{\d z}>0,\ee 
 where (cf. also \eqref{apm})
$$  \tau^{y}(z ,H,B) 
  :=\frac{{\rm Argtanh} \left( A^\al z  \right)}{\al\sqrt{2H}}-\frac{{\rm Argtanh} \left( A^{-\al} z  \right)}{\al\sqrt{2H+4\al\bt^2}}.$$
  Since $A^\al>A^{-\al}$, we find
    $$ \frac{\d \tau^y}{\d z}=\frac{A^\al}{\al\sqrt{2H}(1-(A^\al)^2z^2)}-\frac{A^{-\al}}{\al\sqrt{2H+4\al\bt^2}(1-(A^{-\al})^2z^2)}>$$$$ >\frac{A^{\al}\sqrt{2H+4\al\bt^2}-A^{-\al} \sqrt{2H } }{\al B(1-(A^{-\al})^2z^2)\sqrt{2H+4\al\bt^2}\sqrt{2H }}=$$
 \be=\frac{A^{\al}\sqrt{K^2-(\bt^2-\al B^2)^2}-A^{-\al} \sqrt{K^2-(\bt^2+\al B^2)^2 }}{\al B(1-(A^{-\al})^2z^2)\sqrt{2H+4\al\bt^2}\sqrt{2H }}>0.\label{ine}\ee
 Proving the last inequality in \eqref{ine} therefore boils down to proving the 
following inequality 
$$\sqrt{\frac{K+(\bt^2+\al B^2)}{
  K -(\bt^2+ \al B^2) }}\sqrt{K^2-(\bt^2-\al B^2)^2}>\sqrt{\frac{K+(\bt^2- \al B^2)}{
  K -(\bt^2-\al B^2) }}\sqrt{K^2-(\bt^2+\al B^2)^2 },$$
  which is equivalent to the evident inequality
  \be \frac{1}{{
  K -(\bt^2+ \al B^2) }}>\frac{1}{
  K -(\bt^2- \al B^2) }.\ee

 \end{document}